\documentclass[a4paper,11pt]{article}
\usepackage{aaskaiid}
\usepackage{orcidlink}
\usepackage{rotating}

\title{Measurements of magnetic fields in circumnuclear matter with the SKA telescopes}
\ShortTitle{Magnetic fields in circumnuclear matter}

\author[1,2]{Satoko Sawada-Satoh\orcidlink{0000-0001-7719-274X}}
\ShortName{Sawada-Satoh et al.} 
\author[3,4]{Yuki Kudoh}

\affiliation[1]{Department of Electrical, Electronic and Computer Engineering, Faculty of Engineering, \\ Fukui University of Technology, Gakuen, Fukui-City, Fukui, Japan}
\affiliation[2]{Graduate School of Science, Osaka Metropolitan University,  Sugimoto, Sumiyoshi-ku, Osaka, Japan}
\emailAdd{swdsth@gmail.com}
\affiliation[3]{Astronomical Institute, Tohoku University, Sendai, Miyagi, Japan}
\emailAdd{yuki.kudoh@astr.tohoku.ac.jp}
\affiliation[4]{Information Technology Center, The University of Tokyo, Kashiwa, Chiba, Japan}
\emailAdd{yuki.kudoh@cspp.cc.u-tokyo.ac.jp}

\abstract{
Magnetic fields are thought to regulate the angular momentum transfer in active galactic nuclei (AGNs),  yet their strength and structure in circumnuclear regions remain poorly constrained across spatial scales and gas phases. We present a unified observational framework for probing circumnuclear magnetic fields using complementary diagnostics: direct measurements via the Zeeman effect in H~\textsc{i} absorption and  megamaser emission, and indirect constraints from broadband Faraday rotation of polarized continuum radiation. These approaches provide access to magnetized gas spanning spatial scales from $\sim$100 parsec (pc) circumnuclear disks down to sub-pc regions near supermassive black holes (SMBHs). 
The Square Kilometre Array (SKA) telescopes are expected to revolutionize such investigations through their outstanding sensitivities, wide frequency ranges and high spatial resolutions achievable via very long baseline interferometry (VLBI). 
These capabilities will enable the detection and detailed characterization of weakly polarized emission from magnetized circumnuclear matter, which has remained largely inaccessible with current instruments.
In this chapter, we review previous measurements of magnetic fields in galactic nuclei and discuss  the breakthroughs that SKA observations are expected to provide in elucidating the physical conditions and processes shaping AGN environments.
}

\begin{document}
\maketitle

\section{Introduction}
According to the Unified Scheme of AGNs
\citep[e.g.][]{antonucci93}, 
an AGN consists of 
a SMBH located at the center of a galaxy
and dense gas and dust surrounding it.  
The circumnuclear matter forms distinct structural components depending on spatial scale and physical properties, most notably the accretion disk 
(on sub-pc scales) 
and the torus (a few pc to tens of pc). 
It is widely accepted that 
SMBHs in AGNs grow by accreting gas and dust from their surrounding environment such as accretion disks and tori.
The removal of angular momentum from the accreting gas is a fundamental requirement for mass inflow onto SMBHs in  AGNs. 
Among the various mechanisms proposed, magnetic fields are widely recognized as one of the primary driver of angular momentum transport in circumnuclear matter, through the process of magnetorotational instability \citep[MRI; ][]{balbus91,balbus98}.
MRI occurs in differentially rotating disks with magnetized matter and generates magnetohydrodynamic turbulence.
This turbulence moves angular momentum outward, allowing the matter to slowly move inward toward the SMBH.

Observationally constraining the strength and configuration of magnetic fields within the central tens of pc of AGNs is therefore essential for validating theoretical models of accretion.  
 These regions serve as critical transition zones where material from the host galaxy is funneled toward the SMBH, and where magnetic fields likely regulate the efficiency of angular momentum transport. 
Polarimetric observations of the circumnuclear matter  surrounding  SMBHs can provide direct information about magnetic field structures. 
To date, magnetic fields in the interstellar medium 
have been measured 
in a small number of central regions of external galaxies.  
In this report, we describe past measurements of magnetic fields in the central region of galaxies,  
and discuss how SKA telescopes will provide new insights into magnetic fields in the circumnuclear environment. 
Magnetic fields in circumnuclear regions can be probed using two complementary observational approaches. The Zeeman effect provides a direct measurement of the line-of-sight magnetic field in spectral lines, while Faraday rotation of polarized continuum emission offers an indirect but statistically powerful probe of magnetized plasma. In the following sections, both methods are reviewed, and their synergy in the SKA era is discussed.

\section{Measurement of Magnetic Fields via the Zeeman Effect}

\subsection{Measurement techniques}

One of the most direct methods for probing magnetic fields in astrophysical environments is the detection of the Zeeman effect. 
It refers to the splitting of atomic or molecular spectral lines into right- and left-circularly polarized components (RCP and LCP, respectively) under the influence of an external magnetic field.
These polarized components appear slightly shifted to longer and shorter wavelengths relative to the original line center. 
The frequency shift is proportional to the magnetic field strength, specifically the component along the line of sight (LOS) to the observer.
Such measurements are commonly applied to specific spectral lines, including 21~cm transition of neutral atomic hydrogen (H~\textsc{i}) and molecular lines from species such as OH and CN.
In particular, maser emission lines (e.g., OH, H$_2$O) offer ideal conditions for Zeeman detection due to their narrow intrinsic line widths and high degrees of polarization, which enhance the detectability of subtle frequency shifts.

The difference between the polarized components, 
or the Stokes $V$ spectrum ($V =$ RCP $-$ LCP)  can be approximately given by 
\begin{equation}
V(\nu) = \frac{\mathrm{d}I(\nu)}{\mathrm{d}\nu}\, Z\, B_{\mathrm{los}}
	\label{eq:stokesv}
\end{equation}
where $I$ is the Stokes $I$ spectrum
($I =$ RCP $+$ LCP), 
$\nu$ is the frequency,
$Z$ is the Zeeman splitting factor,
and 
$B_{\rm los}$
is the LOS component of the magnetic field \citep{heiles93}. 

In extragalactic environments, Zeeman measurements are primarily feasible in H~\textsc{i} absorption and in molecular maser emission. These tracers probe different physical scales and gas phases, thereby providing complementary insights into magnetic fields from tens of pc down to sub-pc regions around the central engine.

\subsection{H~\textsc{i} Zeeman measurements}

\begin{figure}[bt]
    \centering
	\includegraphics[width=0.5\columnwidth]{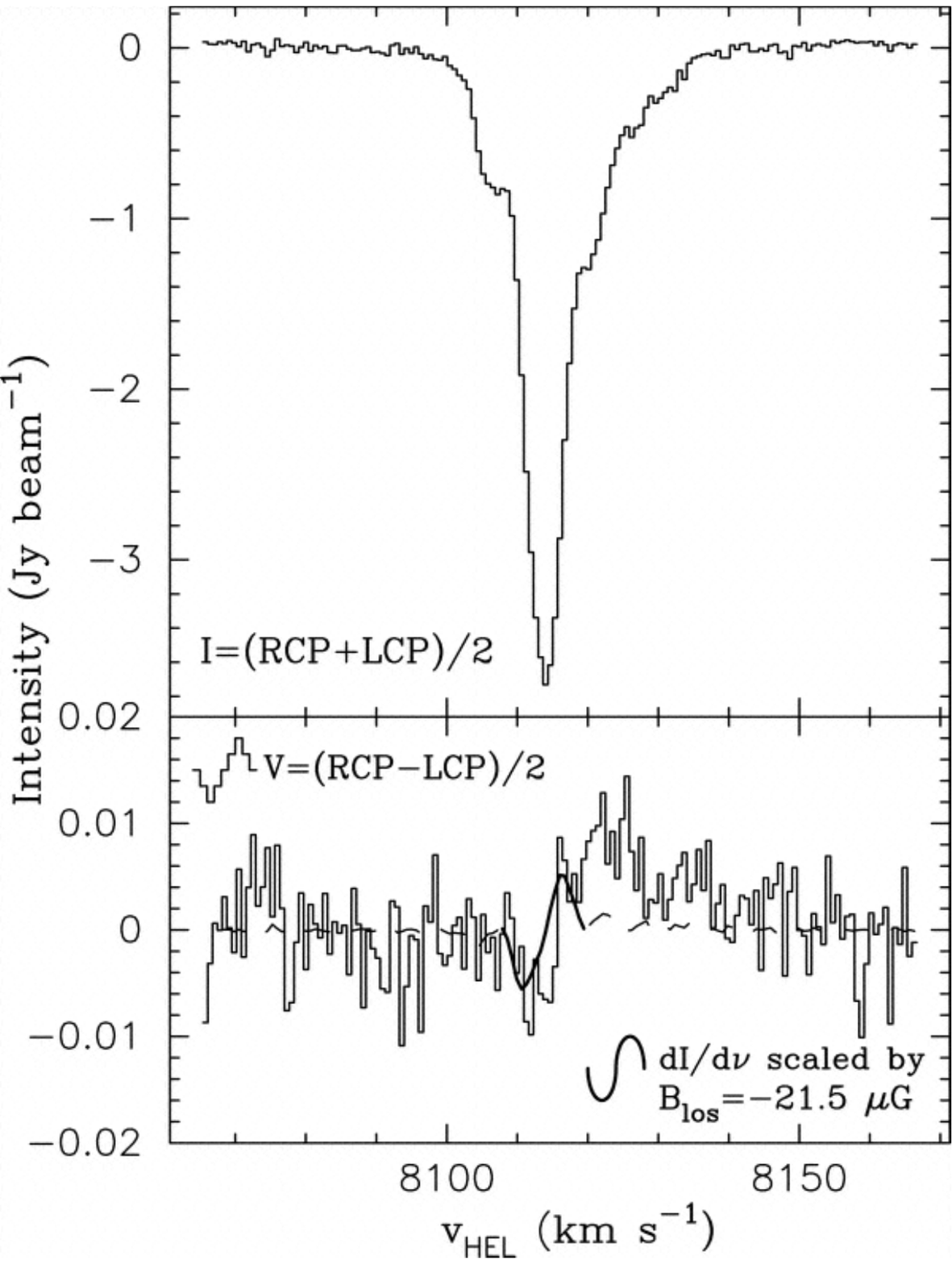}
    \caption{Stokes $I$ (\textit{top}) and $V$ (\textit{bottom}) spectral profiles of the 21 cm 
    H~\textsc{i} absorption in NGC 1275. 
    These panels are exactly the same as Figure 3 in \citet{sarma05}.}
    \label{fig:n1275hi}
\end{figure}

A number of measurements of the magnetic field strengths  
in the interstellar medium 
have been undertaken using the Zeeman effect  
\citep[e.g.,][]{heiles76, troland82, crutcher19}. 
However, 
the detections of extragalactic Zeeman splitting remain rare. 
The first detection of extragalactic Zeeman splitting was reported 
 with the Nan\c{c}ay radio telescope by \citet{kazes91} 
and later confirmed using the Very Large Array (VLA) by \citet{sarma05} 
in an H~\textsc{i} absorption line in the elliptical galaxy NGC 1275 (Perseus A). 
This galaxy has two H~\textsc{i} gas velocity systems, 
one at the systemic velocity of $\sim$5200 km~s$^{-1}$ 
and the other at $\sim$8200 km~s$^{-1}$ 
(i.e., the high velocity cloud). 
\citet{kazes91} and \citet{sarma05} measured the magnetic field 
strengths in the high velocity  H~\textsc{i} cloud, 
yielding LOS magnetic field strengths of $-9.3$ and $-21.5$ $\mu$G, respectively (Figure~\ref{fig:n1275hi}). 
Based on subsequent interferometric and VLBI observations, \citet{morganti23}
argued that the H~\textsc{i} absorption originates from gas associated with a circumnuclear disk on a scale of approximately 100 pc.

Following the detection of the Zeeman splitting in H~\textsc{i} 
absorption of NGC 1275, 
another H~\textsc{i} Zeeman observation toward NGC 5128 (Centaurus A) was conducted, 
which set an upper limit on the LOS magnetic field 
($\sim - 10 \pm 20$ $\mu$G) 
in the region traced by H~\textsc{i} absorption 
toward the nucleus \citep{sarma02}. 
The H~\textsc{i} clouds responsible for the absorption were interpreted to reside in a circumnuclear ring with a radius of about 100 pc. 
To date, no Zeeman splitting has been confirmed in H~\textsc{i}  absorption lines in AGNs on smaller scales.  
Detecting such signatures therefore requires observations with both high sensitivity and high spatial resolution. 



\subsection{Zeeman measurements of megamasers}


Megamasers are extremely luminous astrophysical masers that occur in external galaxies, 
emitting microwave radiation that is millions of times stronger than that of typical Galactic masers. 
They are produced by molecular species such as hydroxyl (OH), water (H$_2$O) and methanol (CH$_3$OH) under specific physical conditions such as high gas densities, elevated temperatures and intense infrared or radio radiation fields. These conditions are often found in the circumnuclear regions of active galaxies or in merging systems with powerful starbursts.

OH megamasers are generally found in luminous and ultraluminous infrared galaxies \citep[e.g.][]{baan89,darling02}. 
They are thought to be excited by dense molecular gas and strong infrared radiation associated with vigorous star formation or galaxy mergers. OH megamasers are generally associated with starburst activity rather than AGN, as their excitation is primarily driven by the intense far-infrared radiation fields produced by dust-heating young stars in merging systems.
\citet{robishaw08}
reported the first clear detections of Zeeman splitting in extragalactic OH megamasers, providing direct measurements of magnetic fields in luminous and ultraluminous infrared galaxies. 
Zeeman splitting was detected in five of eight OH megamaser galaxies observed at 1667 MHz with the Arecibo and Green Bank Telescopes (GBT).
\citet{mcbride15} measured the magnetic fields in the circumnuclear region of the OH megamaser galaxy Arp 220 with the High Sensitivity Array (Arecibo, GBT, the VLBA, the phased VLA), achieving a spatial resolution of about 10 pc.  
This is the first VLBI detection of Zeeman splitting in an extragalactic source. 
Their main results show magnetic field strengths of  $\sim$1--5 mG associated with three individual maser clouds in the center of Arp 220, each of which is roughly pc in size.
Importantly, the measured field strengths and their LOS orientations agree in magnitude and sign with earlier single‐dish Zeeman measurements by 
\citet{robishaw08} 
for features at the same velocities. This agreement validates the use of large‐dish single‐antenna Zeeman observations for OH megamasers.

In contrast, H$_2$O megamasers are typically located in the circumnuclear regions of AGN within $\sim$1 pc. 
In some cases, H$_2$O megamasers have been found to originate in accretion disks at sub-pc radii. 
\citet{modjaz05} carried out the polarimetric observations of the H$_2$O masers in NGC 4258 using the VLA and GBT 
aiming to detect Zeeman-induced  circular polarization and thus measure magnetic fields in the sub-pc accretion disk. 
However, no significant polarization was detected, giving upper limits of about 90 mG for the toroidal and 30 mG for the radial magnetic components.
Because the Zeeman effect of H$_2$O is extremely small 
compared to H~\textsc{i} or OH 
\cite[e.g.,][]{vlemmings02}, 
the detection of Zeeman splitting requires very high spectral resolution ($\ll$ 0.1 km~s$^{-1}$) and excellent signal-to-noise (S/N) ratios.

\subsection{Relevant spectral lines for Zeeman measurements}

The principal spectral lines used for Zeeman splitting measurements are summarized in Table~\ref{tab:lines}, 
together with their rest frequencies and their correspondence with the SKA-Mid frequency bands. Table~\ref{tab:lines} includes both transitions accessible within the current SKA AA4 baseline (Bands~1, 2 and 5) and those requiring future frequency extensions including Bands 3 and 4 (see Section~\ref{sec:zeemanfuture}).

The most important transitions for Zeeman studies are concentrated at frequencies of $\sim$1.4--1.7~GHz, where the H\,{\sc i} 21~cm line and the OH ground-state lines are located. These lines fall within SKA-Mid Band~2 and therefore represent the primary tracers of magnetic fields accessible with the current SKA configuration. Methanol maser transitions at 6.7 and 12.2~GHz are also covered by Band~5, although their Zeeman splitting coefficients are relatively small and remain uncertain.

At higher frequencies, the 22~GHz H$_2$O maser line serves as a unique probe of magnetic fields in sub-pc regions around active galactic nuclei, but lies beyond the upper frequency limit of SKA-Mid. Observations of this transition would require future extensions to higher frequencies.
Band 6 has been proposed as a possible future extension of SKA-Mid but is not part of the currently confirmed SKA AA4 baseline.

In the following, we assess the feasibility of Zeeman measurements based on the currently planned SKA AA4 capabilities and then discuss the additional opportunities offered by future extensions. 


\begin{table*}
\begin{center}
\caption{Zeeman-sensitive spectral lines and their correspondence with SKA-Mid frequency bands.}
\label{tab:lines}

\begin{tabular}{l l c c c}
\hline
Species & Transition & Frequency & SKA-Mid Band$^{a}$ & Zeeman Sensitivity \\
 & & (GHz) &  &  \\
\hline
HI   & 21 cm (F=1--0) & 1.420 & Band 2  & Moderate \\
OH   & 1612 MHz       & 1.612 & Band 2  & High     \\
OH   & 1665 MHz       & 1.665 & Band 2  & High     \\
OH   & 1667 MHz       & 1.667 & Band 2  & High     \\
OH   & 1720 MHz       & 1.720 & Band 2  & High     \\
CH$_3$OH & 6.7 GHz   & 6.668 & Band 5   & Low--uncertain \\
CH$_3$OH & 12.2 GHz  & 12.178 & Band 5  & Low--uncertain \\
H$_2$O   & 22 GHz    & 22.235 & Band 6$^{b}$ & Very low \\
\hline
\end{tabular}
\end{center}
\vspace{1mm}
{\footnotesize
$^{a}$ 
The SKA-Mid frequency coverage includes 
Band 1 (0.35-1.05 GHz), Band 2 (0.95--1.76 GHz), 
and Band 5 (4.60-15.30 GHz) within the AA4 baseline, 
while Band 3 (1.65-3.05 GHz) and Band 4 (2.80-5.18 GHz) 
are planned for future stages. 
\\
$^{b}$ Band 6 refers to a proposed (not yet confirmed) extension of SKA-Mid to frequencies above $\sim$15 GHz, beyond the planned AA4 scope.
}

\end{table*}

\subsection{
Observational prospects with the SKA AA4 baseline
}

In this section, we focus exclusively on the capabilities within the currently planned SKA AA4 baseline, which includes Bands 1, 2, and 5 (0.35--15.3 GHz), while Bands 3 and 4 are not part of the AA4 configuration. The full frequency range of SKA-Mid spans Bands 1--5, but only a subset is available in the current baseline. 
The frequency coverage overlaps with the standard observing bands of existing global VLBI facilities, including the LBA \citep{edwards15,lovell13}, VLBA \citep{napier94}, EVN \citep{paragi15}, and EAVN  \citep{akiyama22}. Together, these facilities provide global coverage in standard radio bands such as L, S, C, and X ($\sim$1--15 GHz), thereby enabling joint observations with SKA-Mid as part of VLBI networks.

Within this frequency range, Zeeman measurements are primarily enabled through H\,{\sc i} absorption against bright radio continuum sources. Such observations require high sensitivity and angular resolution, which can be achieved through VLBI  including SKA-Mid. By participating in VLBI networks, SKA-Mid can resolve structures on scales of a few pc, allowing detailed studies of circumnuclear gas in active galactic nuclei.

\subsubsection{Target sources}

Observed flux density plays a key role in assessing the feasibility of Zeeman measurements, making it essential to prioritize the brightest available sources. Suitable targets for H\,{\sc i} Zeeman observations should be associated with a strong and compact radio continuum background, typically with flux densities of at least several hundred mJy. In addition, clearly detected H\,{\sc i} absorption with sufficient optical depth is required, preferably with relatively simple and narrow velocity components to minimize blending effects that can obscure the Zeeman signature.

Compact structure on VLBI scales is also crucial to isolate circumnuclear regions on scales of $\sim$100 pc. Nearby radio galaxies such as NGC~5128 \citep[e.g.,][]{morganti08} and NGC~4261 \citep[e.g.,][]{vanlangevelde00} satisfy these conditions and are accessible from the southern hemisphere. NGC~1275 can also be considered a viable candidate despite its relatively high declination, given its strong continuum emission and previously detected H\,{\sc i} absorption. These systems exhibit bright ($>1$~Jy) and compact radio continuum emission at 1.4~GHz, making them well suited for sensitive Zeeman measurements.

\subsubsection{Sensitivity}

As a sensitivity estimate for VLBI observations using the phased SKA-Mid array \citep[SEFD $\sim$2~Jy at 1.4~GHz][] {braun19}, we consider baselines with the 64-m Parkes telescope \citep[SEFD $\sim$30~Jy][]{hobbs20} and the 26-m HartRAO antenna (SEFD $\sim$350~Jy). For a spectral resolution of 0.21~kHz and an integration time of 10~hours, the resulting baseline sensitivities are approximately 2.3~mJy for SKA-Mid--Parkes and 7.8~mJy for SKA-Mid--HartRAO.

The detectability of H\,{\sc i} Zeeman splitting can be evaluated by comparing these noise levels with the expected Stokes $V$ signal. Observations of NGC~1275 suggest that the Stokes $V$ amplitude typically ranges from a few mJy up to $\sim$10~mJy \citep{sarma05}. For the SKA-Mid--Parkes baseline, signals at the higher end of this range should be detectable with moderate significance, whereas weaker signals would likely remain marginal without longer integration times. In contrast, for SKA-Mid--HartRAO, the expected noise level is comparable to or larger than the signal amplitude, limiting detections to only the strongest cases.

The corresponding Zeeman splitting scale is approximately 75~$\mu$G for a spectral resolution of 0.21~kHz. In practice, the minimum detectable magnetic field strength depends on the S/N ratio, and values of a few tens of $\mu$G may be achievable for sufficiently strong sources. An optimal channel width of 100--300~Hz delivers a good balance between spectral resolution and sensitivity, making a resolution of 0.21~kHz well suited for H\,{\sc i} Zeeman observations within the SKA AA4 framework.

\subsection{Additional capabilities with future upgrades}
\label{sec:zeemanfuture}

We now consider additional capabilities enabled by potential future upgrades beyond the SKA AA4 baseline.
Future extensions of the SKA to higher frequencies would significantly expand the range of accessible Zeeman diagnostics. In particular, the 22~GHz H$_2$O maser line offers a unique probe of magnetic fields in sub-pc regions around SMBHs, tracing the innermost parts of accretion disks in active galactic nuclei. 
However, this capability depends on a proposed extension of SKA-Mid to frequencies above $\sim$15~GHz (hereafter referred to as ``proposed Band~6''), which is not currently included in the SKA AA4 baseline. Zeeman measurements of H$_2$O masers therefore represent an important science case for future upgrades of the SKA system.

\subsubsection{Target sources}

Nearly 200 galaxies are known to host 22~GHz H$_2$O maser emission, of which about 150 are associated with AGNs \citep[e.g.,][]{braatz18}. The most promising targets for Zeeman studies are those with flux densities above $\sim$1~Jy, such as the Circinus galaxy and NGC~4945. These nearby southern sources exhibit strong maser emission, resulting in high S/N observations. Sources with intermediate flux densities (0.1--1~Jy) are also viable targets, provided sufficiently long integration times are used.

Expanding the number of observed sources will be essential for establishing statistically meaningful constraints on magnetic fields in circumnuclear environments. At present, Zeeman measurements in AGN disks remain limited to a small number of objects, preventing robust conclusions about typical magnetic field strengths and their dependence on nuclear properties.

\subsubsection{Sensitivity}

Assuming that the sensitivity of the proposed Band~6 receiver is comparable to that of Band~5, the thermal noise level for SKA-Mid is estimated to be $\sim$1.7~mJy~beam$^{-1}$ for an integration time of 4~hours at a spectral resolution of 0.21~kHz. At 22~GHz, this corresponds to a velocity resolution of approximately 2.9~m~s$^{-1}$.

For a source with a flux density of 1~Jy and a fractional circular polarization of 1\%, the expected Stokes $V$ signal is $\sim$10~mJy, yielding a S/N ratio of about 6. This indicates that Zeeman measurements of H$_2$O maser emission would be feasible for sufficiently bright sources. Longer integration times would further improve sensitivity and enable observations of weaker systems.

The inclusion of higher-frequency capabilities would therefore open a new regime for magnetic field studies.
It would complement lower-frequency Zeeman tracers and provide a unified view across spatial scales of magnetized environments near SMBHs. 
In particular, H$_2$O maser observations provide a unique and essential probe of magnetic fields on sub-pc scales, which cannot be accessed within the current SKA AA4 frequency range.

Taken together, these considerations highlight the importance of extending the accessible frequency range. Access to the 22 GHz H$_2$O maser line is essential for probing magnetic fields on sub-pc scales. Without such high-frequency capability, the innermost regions of AGN accretion disks remain largely inaccessible. This strongly motivates the inclusion of a high-frequency extension such as the proposed Band 6.

\section{Measurements of Magnetic Fields via  Rotation Measure}

In this section, we first summarize existing RM measurements (Sections 3.1--3.2), and then present the science case for the SKA based on (i) the currently planned AA4 capabilities and (ii) potential future upgrades.

\subsection{Measurement techniques}

The rotation measure (RM) quantifies the Faraday rotation effect experienced by polarized emission as it travels through a magnetized plasma. 
The rotation of the polarization plane in the magnetized plasma is given by 
\begin{equation}
    \textrm{RM} = 0.81 \int n_e B_{\parallel} dl, 
    \label{eq:rm}
\end{equation}
where RM is the rotation measure in rad m$^{-2}$, 
$n_e$ is the electron density in cm$^{-3}$, 
$B_{\parallel}$ is magnetic field in $\mu$G, 
$l$ is the path length in pc along the LOS from the observer \citep[e.g.][]{spitzer78}.  
RM directly traces the LOS magnetic field weighted by the electron density.
Measuring RM across different frequencies allows us to infer the strength, direction and structure of magnetic fields.

\subsection{The past measurement}

\begin{figure}[h]
    \centering
	\includegraphics[width=0.76\columnwidth]{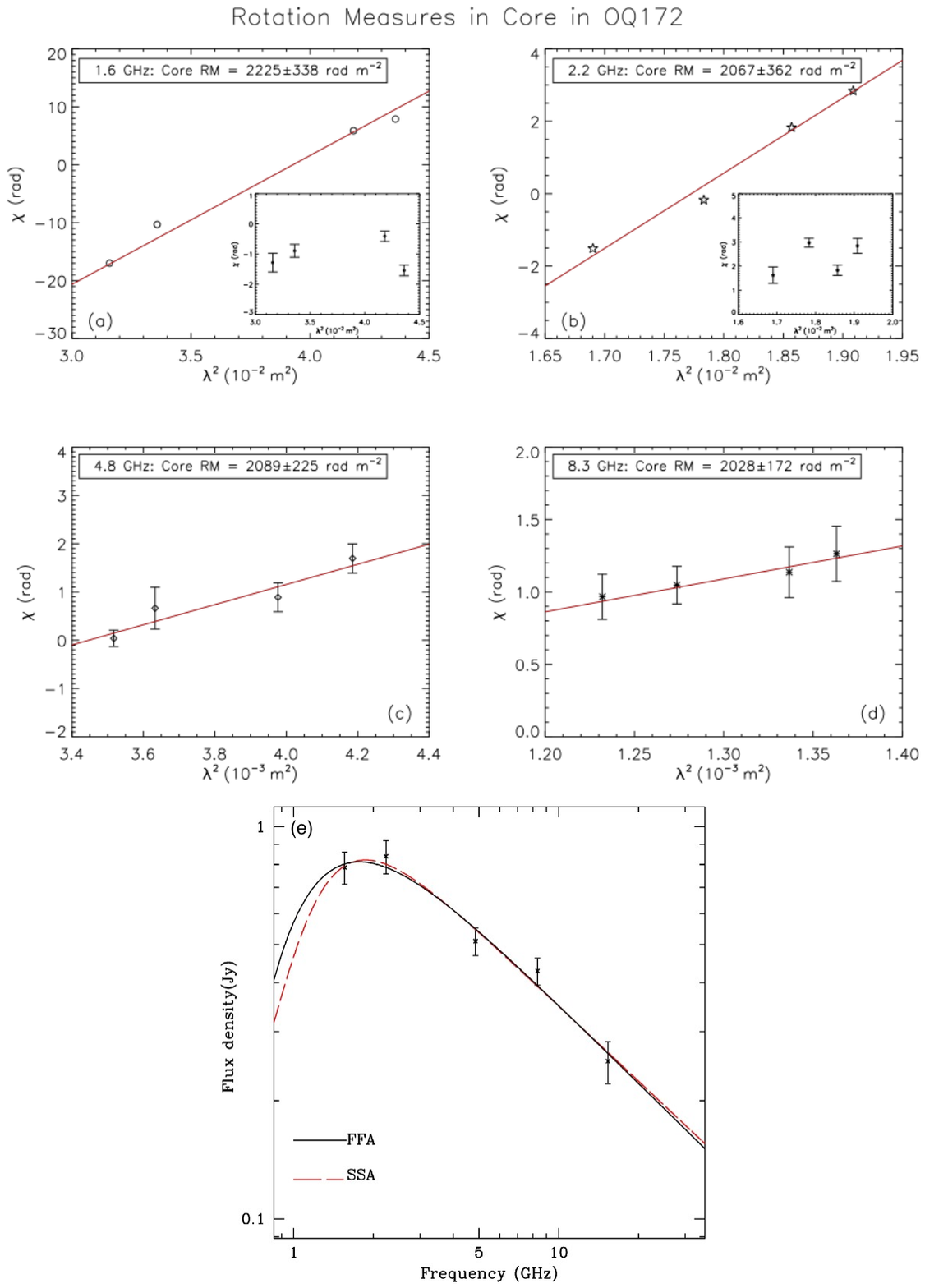}
    \caption{
    RM for the GPS quasar OQ 172 in the core component 
    at (a) 1.6, (b) 2.2, (c) 4.8 and (d) 8.3 GHz. 
    These panels are adapted in part from Figure 4 in \citet{liu17}.
    Spectrum of the core region component 
    at 1.6, 2.2, 4.8, 8.3 and 15.3 GHz is shown in (e). 
    The black solid line and red dashed line indicate the fitted spectra for FFA and SSA models, respectively.
    The two models are not easily distinguished due to the lack of low-frequency observing data. 
    Thus, the value of $n_e$ was estimated by assuming the FFA model.  
    This panel is exactly the same as Figure 7 in \citet{liu17}.}
    \label{fig:oq172}
\end{figure}

\citet{liu17} performed multi-frequency VLBI polarimetric observations of the GHz-peaked spectrum (GPS) quasar OQ~172 and determined RMs over frequencies ranging from 1.6 to 15.3 GHz.
The derived RM is $\sim$2000 rad~m$^{-2}$ in the core component. 
GPS sources are characterized by convex radio spectra that peak at a turnover frequency in the GHz range 
\citep{odea91}. 
At low frequencies below the turnover, the flux density decreases due to absorption processes that make the emitting region optically thick. Two main mechanisms are considered responsible for this behavior: synchrotron self-absorption (SSA) and free-free absorption (FFA). 
To determine which absorption process dominates, measurements at frequencies below the turnover are essential.
Assuming that the shape of the continuum spectrum is caused by FFA due to dense plasma in the core region (Fig.~\ref{fig:oq172}), \citet{liu17} derived $n_e$  from the relationship between $n_e$ and the FFA coefficient $\tau_{f}$ : 
\begin{equation}
    \tau_{f} = 0.08235 ~ \nu^{-2.1} ~ T_{e}^{-1.35} 
    \int n_{e}^2 dl, 
\end{equation}
where $\nu$ is the frequency in GHz, 
$T_{e}$ is the electron temperature in K, 
the electron density $n_e$ in cm$^{-3}$ and
the path-length integral $l$ in pc. 
By combining the independently obtained RM and $n_e$ values through equation~\ref{eq:rm}, $B_{\parallel}$ was thereby estimated to be 2.9 mG.

However, the uncertainty in distinguishing between FFA and SSA models highlights the need for broader frequency coverage and improved sensitivity at low frequencies. The SKA, with its combination of SKA-Low and SKA-Mid, is uniquely positioned to resolve this ambiguity and to provide robust constraints on electron density and magnetic field strength in GPS sources.

\subsection{
Observational prospects with the SKA AA4 baseline
}

Within the SKA AA4 configuration, rotation measure (RM) studies will benefit from the combination of SKA-Low and SKA-Mid, providing wide frequency coverage and high sensitivity. This broadband capability is essential for accurately characterizing Faraday rotation and disentangling absorption processes in the circumnuclear regions of active galactic nuclei.

\subsubsection{Frequency coverage}

A principal objective of RM observations is to constrain free-free absorption (FFA) affecting synchrotron emission from AGN cores. This requires well-calibrated broadband continuum spectra spanning $\sim$0.1--10~GHz. SKA-Low provides access to the low-frequency regime where FFA signatures, such as spectral turnovers and curvature, are most pronounced. SKA-Mid, on the other hand, establishes the intrinsic synchrotron spectrum at higher frequencies, serving as a reference for spectral modeling.

The combination of SKA-Low and SKA-Mid in VLBI observations is therefore critical for distinguishing between synchrotron self-absorption (SSA) and FFA. In addition, wide wavelength-squared ($\lambda^2$) coverage improves the accuracy of RM measurements and reveals complex Faraday structures, providing insight into the magnetized plasma surrounding AGNs.

\subsubsection{Target sources}

In the southern hemisphere, several well-defined samples of compact radio sources are available for RM studies. The Australia Telescope 20~GHz Survey provides a catalog of approximately 5800 sources, including several hundred GHz-peaked spectrum (GPS) and high-frequency peaker (HFP) candidates identified through multi-frequency observations. Additional samples include bright compact steep spectrum (CSS) and GPS sources compiled by \citet{randall11}, as well as the Parkes half-Jansky sample \citet{snellen02}.

These surveys indicate that several tens of bright ($\gtrsim$1~Jy) GPS sources are available in the southern sky. Such sources are ideal targets for RM studies because their compact structure and strong continuum emission allow high S/N polarization measurements and detailed investigation of circumnuclear environments.

\subsubsection{Sensitivity}

Assuming an SEFD of $\sim$3~Jy for SKA-Mid and baseline sensitivities of order tens of $\mu$Jy for SKA-VLBI observations, high-precision RM measurements are achievable. For a source with a correlated flux density of 0.1~Jy and a fractional polarization of 0.1--1\%, the expected polarized flux density is 0.1--1~mJy, corresponding to S/N ratios of $\sim$10--100.

Under these conditions, the uncertainty in polarization angle can be reduced to $\sim$0.3$^\circ$ -- $3^\circ$, leading, in principle, to an RM precision on the order of $10^{-2}$~rad~m$^{-2}$ when using the $\lambda^2$ coverage of SKA-Mid Band~1. This level of precision would enable detailed studies of magnetic field strength and structure in the circumnuclear regions of AGNs within the SKA AA4 framework.

\subsection{
Additional prospects with future SKA extensions
}

We now extend the discussion beyond the SKA AA4 baseline and assess how future upgrades expand the science case relative to the currently planned capabilities. Improvements in sensitivity, frequency coverage, and VLBI capabilities will further strengthen RM studies. In particular, increased sensitivity will allow the inclusion of fainter and more distant sources, significantly expanding the available sample size for statistical analyses.
At present, RM measurements with VLBI are limited to relatively bright sources. Enhanced sensitivity in future SKA configurations will make it possible to probe a much larger population of AGNs, including sources with weaker polarized emission that are currently inaccessible. This expansion is crucial for establishing robust correlations between magnetic field properties and AGN characteristics.

In addition, broader frequency coverage would further improve the determination of RM by increasing the available $\lambda^2$ range. This would allow more precise measurements and better characterization of complex Faraday structures, including multiple RM components and depolarization effects.
Enhanced VLBI capabilities, including improved baseline coverage and sensitivity, will also allow higher-fidelity imaging of polarized emission. This will provide more detailed information on the spatial distribution of magnetic fields and their relationship to AGN jets and circumnuclear gas.
These advances will transform RM studies from investigations of individual objects into systematic surveys, leading to statistically meaningful constraints on magnetic field properties across a wide range of AGN environments.

\section{The Impact of SKA}


The impact of the SKA lies in its ability to unify multiple magnetic field diagnostics across a wide range of spatial scales. By combining Zeeman measurements of spectral lines with broadband Faraday rotation studies of continuum emission, the SKA will provide a coherent, multi-scale view of magnetic fields from $\sim$100 pc circumnuclear disks down to sub-pc accretion regions near SMBHs.
Its key strengths—high sensitivity, broad frequency coverage, and VLBI-based resolution-allow the detection and characterization of weakly polarized emission from circumnuclear matter. These capabilities support detailed spectral modeling, including discrimination between synchrotron self-absorption and free-free absorption, and provide constraints on physical conditions such as electron density, temperature, and magnetic field strength.
In addition, the SKA beyond AA4 is expected to significantly increase the number of accessible targets. Current VLBI studies of H~\textsc{i} absorption and H$_2$O megamasers are limited to a few dozen galaxies. Improved sensitivity will extend these observations to a much larger sample, transforming magnetic field studies from individual detections into systematic and statistically robust investigations.
By combining the complementary capabilities of SKA-Low, SKA-Mid, and SKA-VLBI, future observations will bridge existing gaps in sensitivity, spatial resolution and frequency coverage, leading to a more comprehensive understanding of the role of magnetic fields in AGN accretion and feedback.











\bibliographystyle{abbrvnat-maxbibnames4}


\bibliography{reference} 

\end{document}